\input{aipcheck}

\documentclass[
    ,final            
  ]
  {aipproc}

\layoutstyle{6x9}

\begin{document}

\title 
      [Neutrino Cooled disk in GRBs]
      {Neutrino Cooled disk in GRB central engine}

\classification{97.10.Gz, 97.60.Lf, 98.70.Rz}
\keywords{accretion, accretion discs, black hole physics, neutrinos, gamma-rays: bursts}

\author{A. Janiuk}{
  address={Copernicus Astronomical Center, Bartycka 18, 00-716 Warsaw, Poland},
}

\author{Y. Yuan}{
  address={ Center for Astrophysics, University of Science and Technology of China, Hefei, Anhui 230026, P.R. China},
}

\author{R. Perna}{
  address={Department of Astrophysical and Planetary Sciences, University of Colorado at Boulder, 440 UCB, Boulder, CO 80309, USA\\
and Princeton University, 4 Ivy Lane, Princeton, NJ 08542, USA},
}

\author{T. Di Matteo}{
  address={Carnegie Mellon University, Department of Physics, 5000 Forbes Ave., Pittsburgh, PA 15213, USA},
}

\begin{abstract}

At the extreme densities and temperatures typical of the central
engine of GRBs, the accreting torus is cooled mainly by advection and
by neutrino emission. The latter process is dominated by electron and
positron capture onto nucleons ($\beta$ reactions). We calculate the
reaction rates and the nuclear composition of matter, assuming that
the torus consists of helium, eletron-positron pairs, free neutrons
and protons.  After determining the equation of state and solving for
the disk structure for a given initial accretion rate, we subsequently
follow its time evolution. We find that, for accretion rates of the
order of $10 M_{\odot}$/s, likely typical for the early stages of the
accretion event, the disk becomes unstable, giving rise to variable
energy output. This instability may play an important role
for producing internal shocks.

\end{abstract}

\maketitle

\section{Introduction}

The enormous power that is released
during the gamma-ray burst explosion indicates that a 
relativistic phenomenon must be involved in creating GRBs (\cite{NPP:92}).
Whether GRBs are the end-result of a compact binary merger (e.g. NS-NS or NS-BH), or
the collapse of a massive star, a dense, a hot torus likely forms around
a newly born black hole (\cite{Wittetal:94}).

The steady-state model of a hyperaccreting torus optically thin to neutrinos 
was proposed  e.g. in \cite{PWF:99},  \cite{KM:02}, and the calculations were
performed for accretion rates $\dot M \le 1 M_{\odot}$/s.
The effects of neutrino opacities and trapping were introduced in \cite{DPN:02}.

Since the accretion process is extremely rapid and hence a transient event, it is
best studied by means of a time-dependent disk model
(\cite{Janiuketal:04}). The intial accretion rates in the torus
can be as high as 10 $M_{\odot}$/s, and then rapidly decrease with
time.
The time-dependent disk model can be regarded as complementary to the
studies of binary mergers or collapsars, for which the 2D numerical
simulations were presented by e.g. \cite{McFadWoo:99},
\cite{Rossetal:04}.
Here we present new results, based on the detailed calculations of the equation of state (EOS) and  the
chemical  composition of the nuclear matter.
\section{Model}

For a given accretion rate, we determine the structure of the disk
from the 'steady-state' equations by imposing the balance between viscous heating
($\alpha$-prescription) and cooling due to advection, radiation and
neutrino emission.
The neutrino emission mechanisms are:
(i) Electron - positron capture on nucleons and  $\beta$-decay. These reactions are:
~~ $p + e^{-} \to n + \nu_{e}$, 
~~ $n + e^{+} \to p + \bar\nu_{e}$,
~~ $n \to p + e^{-} + \bar\nu_{e}$
(ii) Annihilation of electron-positron  pairs:
$e^{-}+e^{+}\to \nu_{i}+\bar\nu_{i}$
(iii) Nucleon  Bremsstrahlung:
$n+n \to n+n+\nu_{i}+\bar\nu_{i}$
(iv) Plasmon decay due to the interaction with electron gas:
$\tilde \gamma \to \nu_{e}+\bar\nu_{e}$.

Each of the above neutrino emission process has the reverse process, that 
is the source of neutrino absorption.
In addition, the free escape of neutrinos from the disc is limited by scattering.
Therefore we calculate the absorptive and scattering 
optical depths for neutrinos, and include in the calculations  concerning a non-zero neutrino 
pressure and entropy components.

The equation of state is calculated numerically for a given
temperature and density, which is subsequently iterated in a grid of
radii. We assume that the torus consists of helium, electron-positron
pairs, free neutrons and protons.  The chemical potentials of the
species are calculated from the chemical equilibrium condition.  For a
given baryon number density, $n_{b}=(n_{p}+n_{n})/X_{nuc}$, 
the chemical potentials of neutrons, protons and electrons are 
calculated from the transition reaction rates between neutrons and
protons, using the conditions for the conservation of
baryons and charge neutrality.
%

Having computed the initial state we allow the
density and temperature to vary with time. We solve the time-dependent
conservation equations for mass, angular momentum and energy.
The advection term is included in the energy equation via the radial derivatives. 

\section{Results}

For sufficiently large accretion rates, $\dot M \ge 1 M_{\odot}$/s
neutrino trapping limits the neutrino cooling rate and the central
regions of the disk become advection-dominated.  The temperature
profile (see Fig. 1) for 1 $M_{\odot}$/s is similar to that from
\cite{PWF:99} and \cite{DPN:02} but the density profile is steeper than found in previous models.
For 10 $M_{\odot}$/s we found a distinct branch of solutions that is
unstable due to the helium photodisintegration, which is intrinsically 
incorporated in our EOS calculations. 

\begin{figure}
\includegraphics[height=.3\textheight]{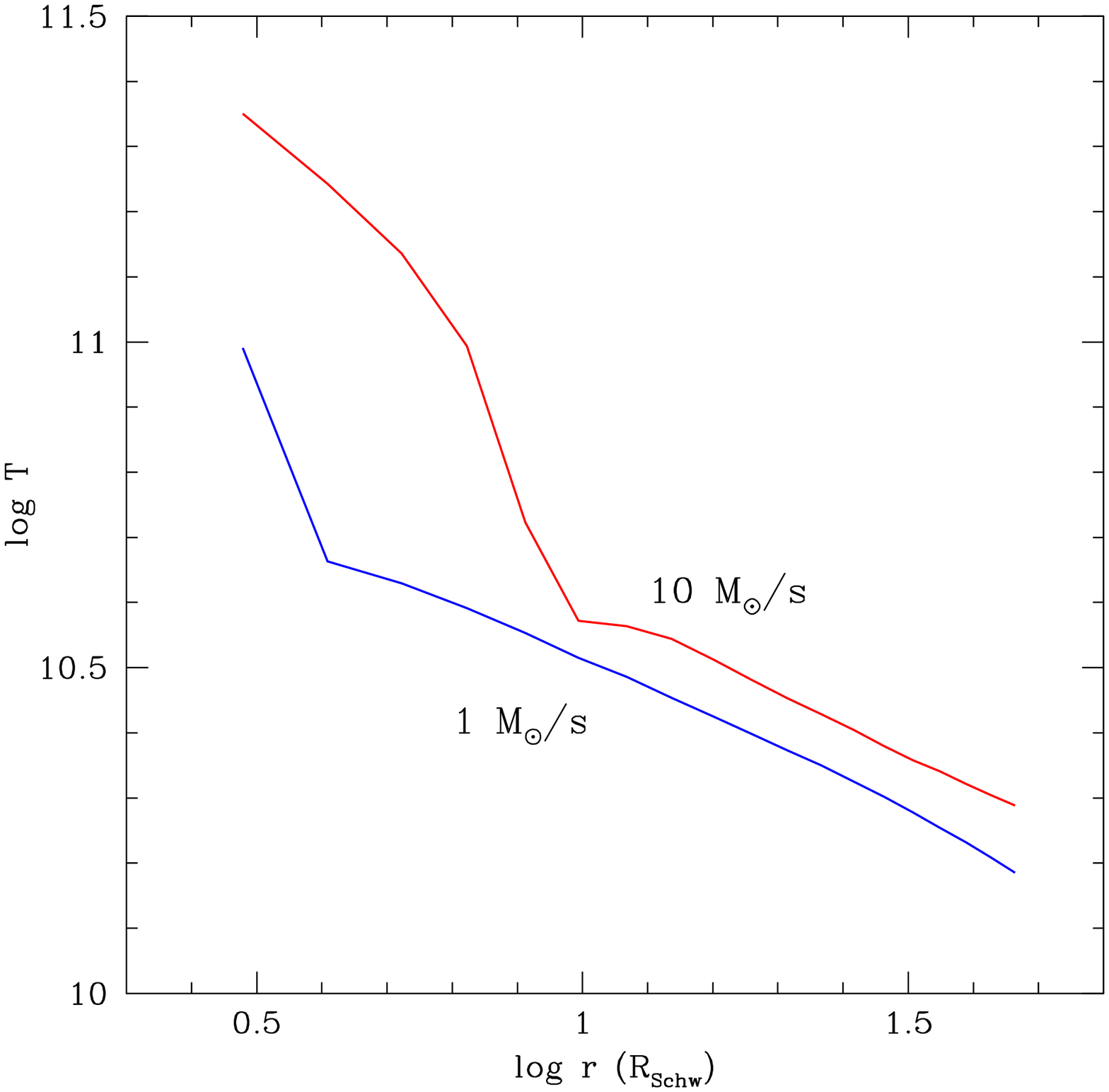}
\includegraphics[height=.3\textheight]{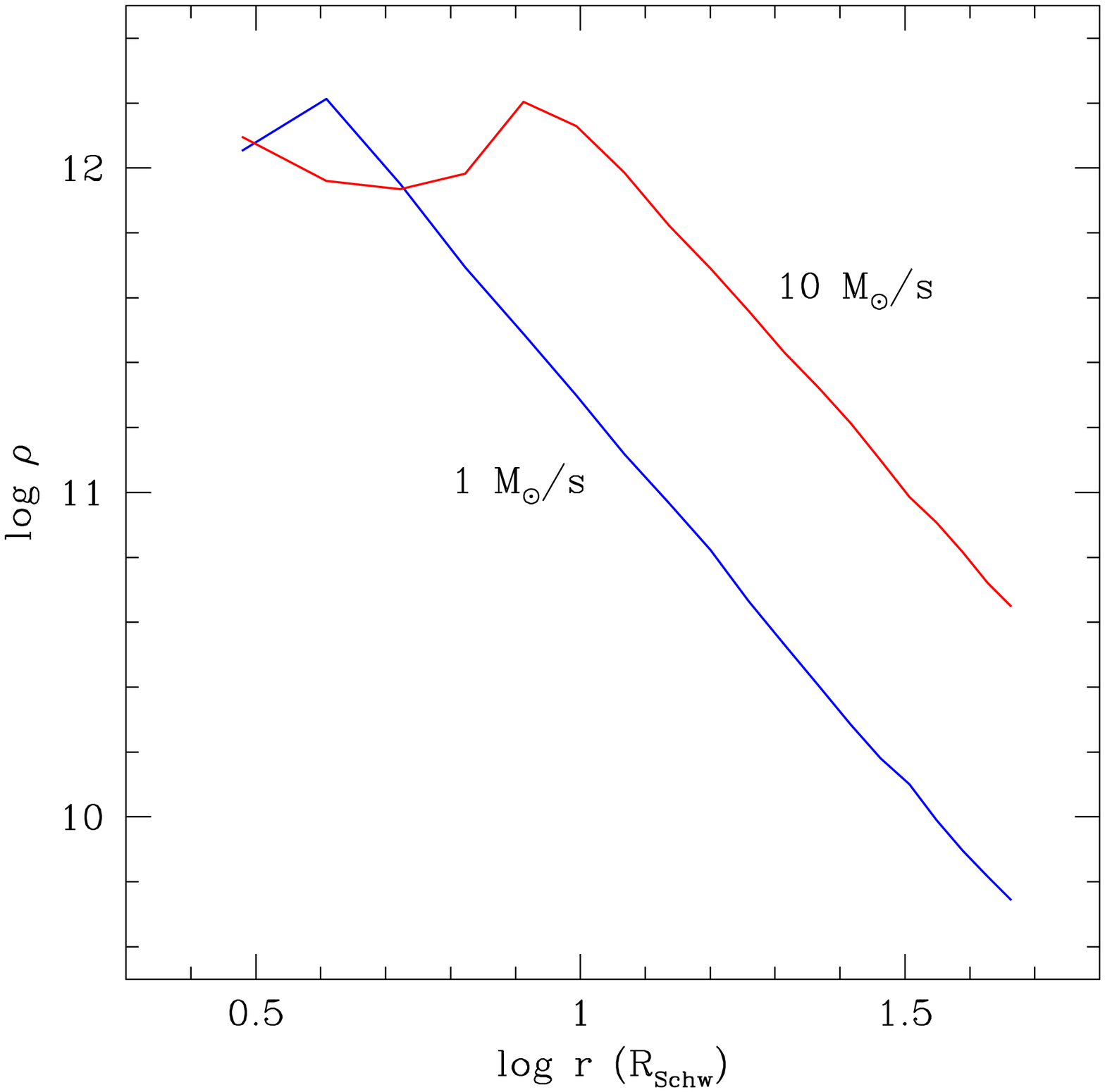}
  \caption{Density and temperature profiles in the disk, for accretion
  rates of 10 $M_{\odot}$/s (upper line) and 1 $M_{\odot}$/s (bottom
  line)}
\end{figure}

{\bf Electron fraction}

\noindent
This is defined as:
$Y_{e} = {n_{e^{-}} - n_{e^{+}} \over  n_{b}}$.
For a given the temperature, the electron fraction changes with 
increasing baryon number density (Fig. 2). At  high densities, $Y_{e}$ roughly
equals 0.5, because the torus consists of plenty of ionized helium and
some electrons 
(the ratio of photon to baryon is small).
As the density decreases,
helium starts to dissolve into free neutrons and protons. After
that, the electron fraction begins to adjust to satisfy the
beta-equilibrium in the gas.  For high temperatures the positrons
appear, as the electrons become non-degenerate; positron capture by
neutrons again increases $Y_{e}$.

\begin{figure}
\includegraphics[height=.3\textheight]{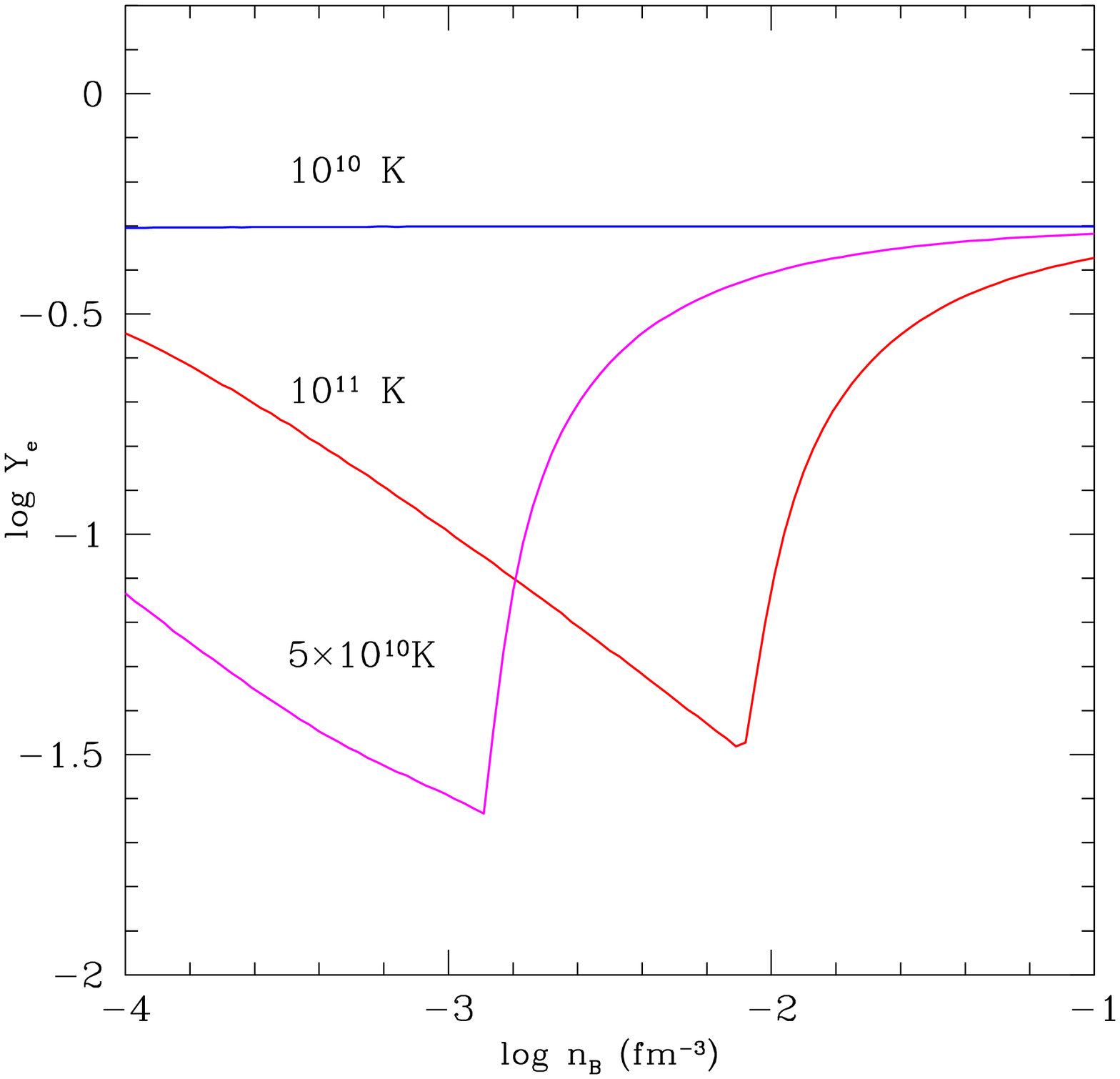}
\includegraphics[height=.3\textheight]{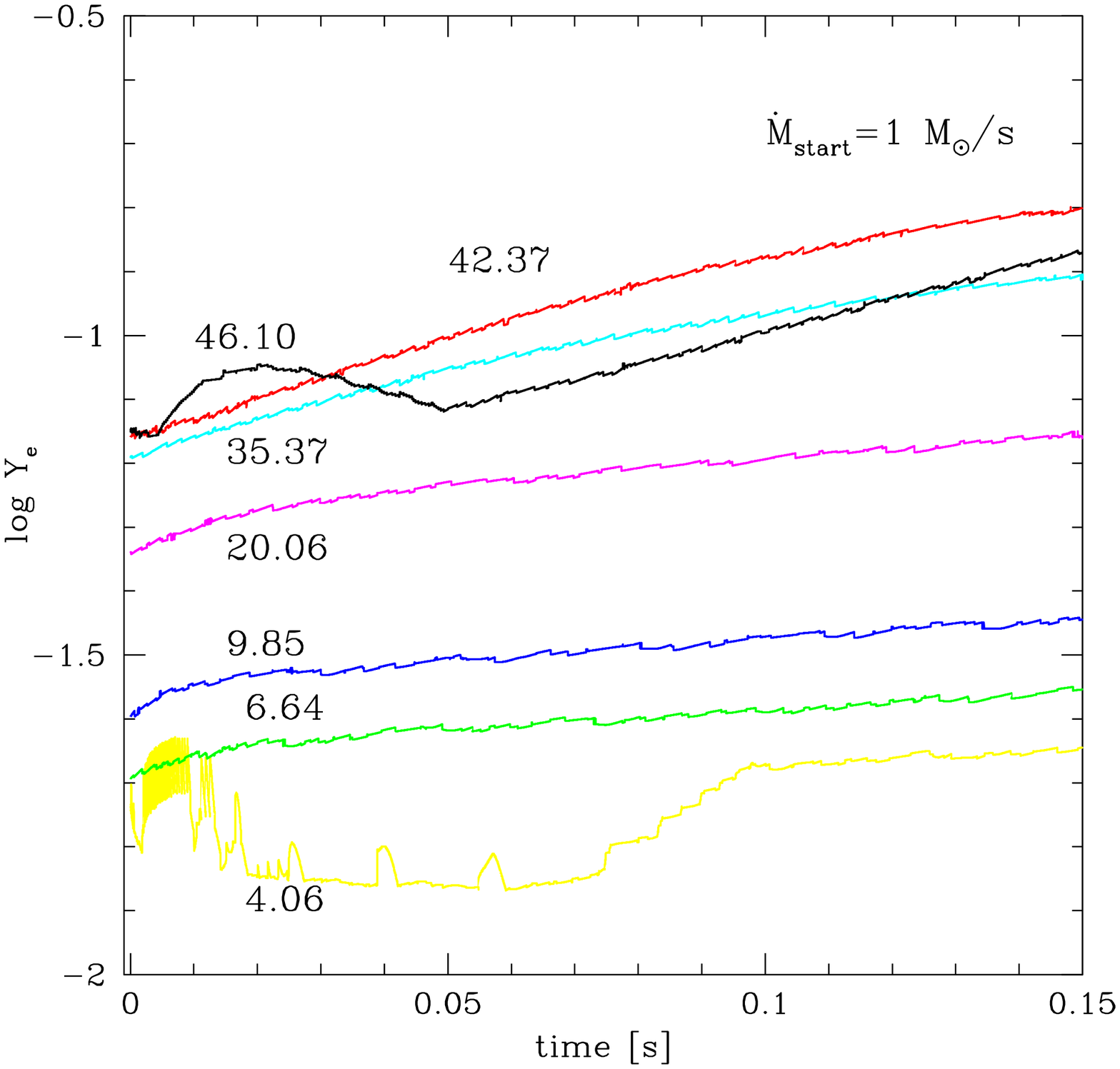}
  \caption{Left: Electron fraction as a function of baryon number density, for 3 various temperatures: $10^{10}$, $5\times 10^{10}$ and $10^{11}$.
Right: Electron fraction as a function of time, at several radii in the accretion disk:
4.06, 6.64, 9.85, 20.06, 35.37, 42.37 and 46.10  $R_{Schw}$.}
\end{figure}

In the time-dependent calculations, initially $Y_{e}$ remains small in the 
innermost disc. Then it starts to increase with time, due to the disk evolution 
and gradual drop in accretion rate, density and temperature.

{\bf Stability of the disk}

\begin{figure}
\includegraphics[height=.3\textheight]{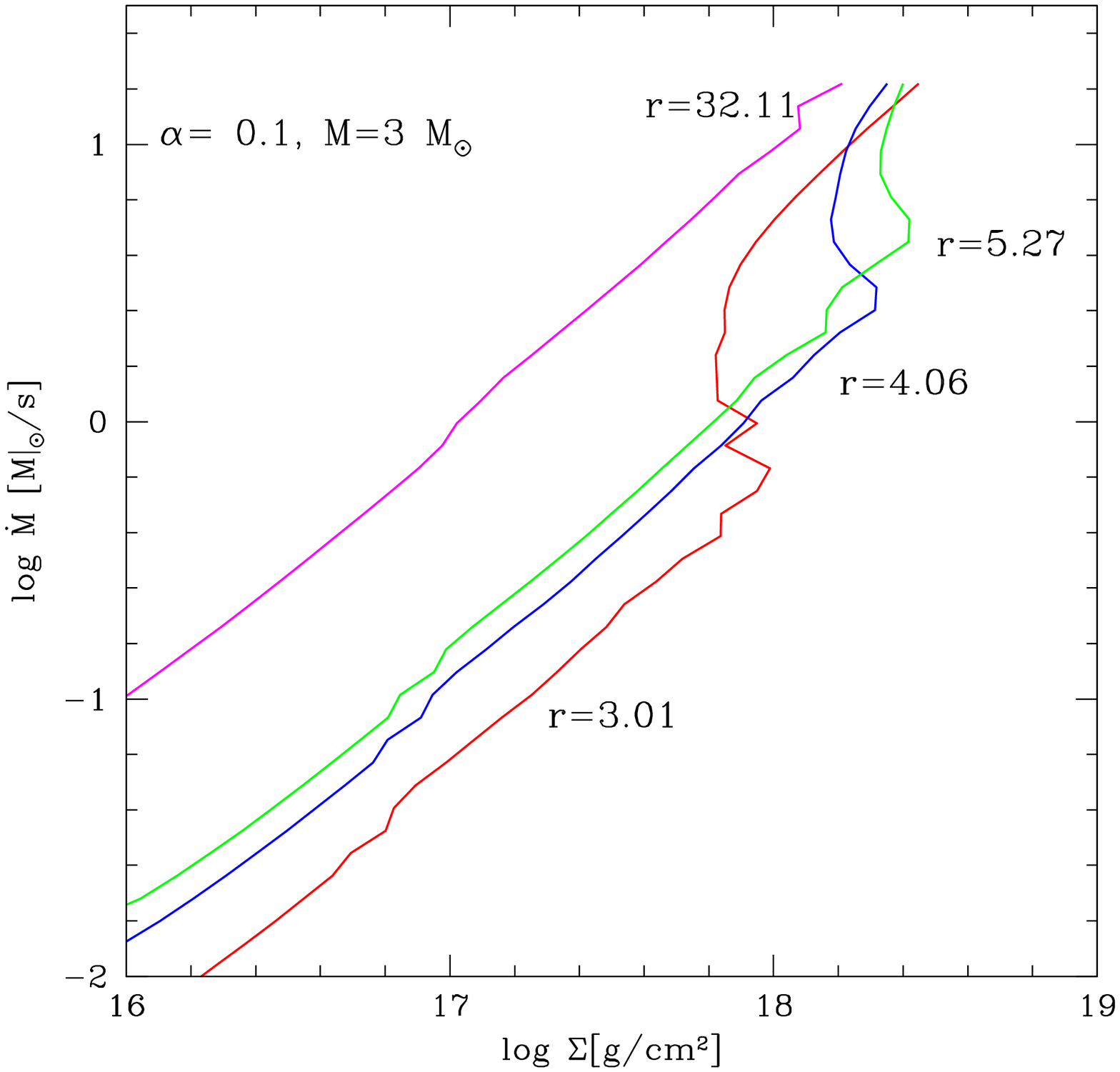}
\includegraphics[height=.3\textheight]{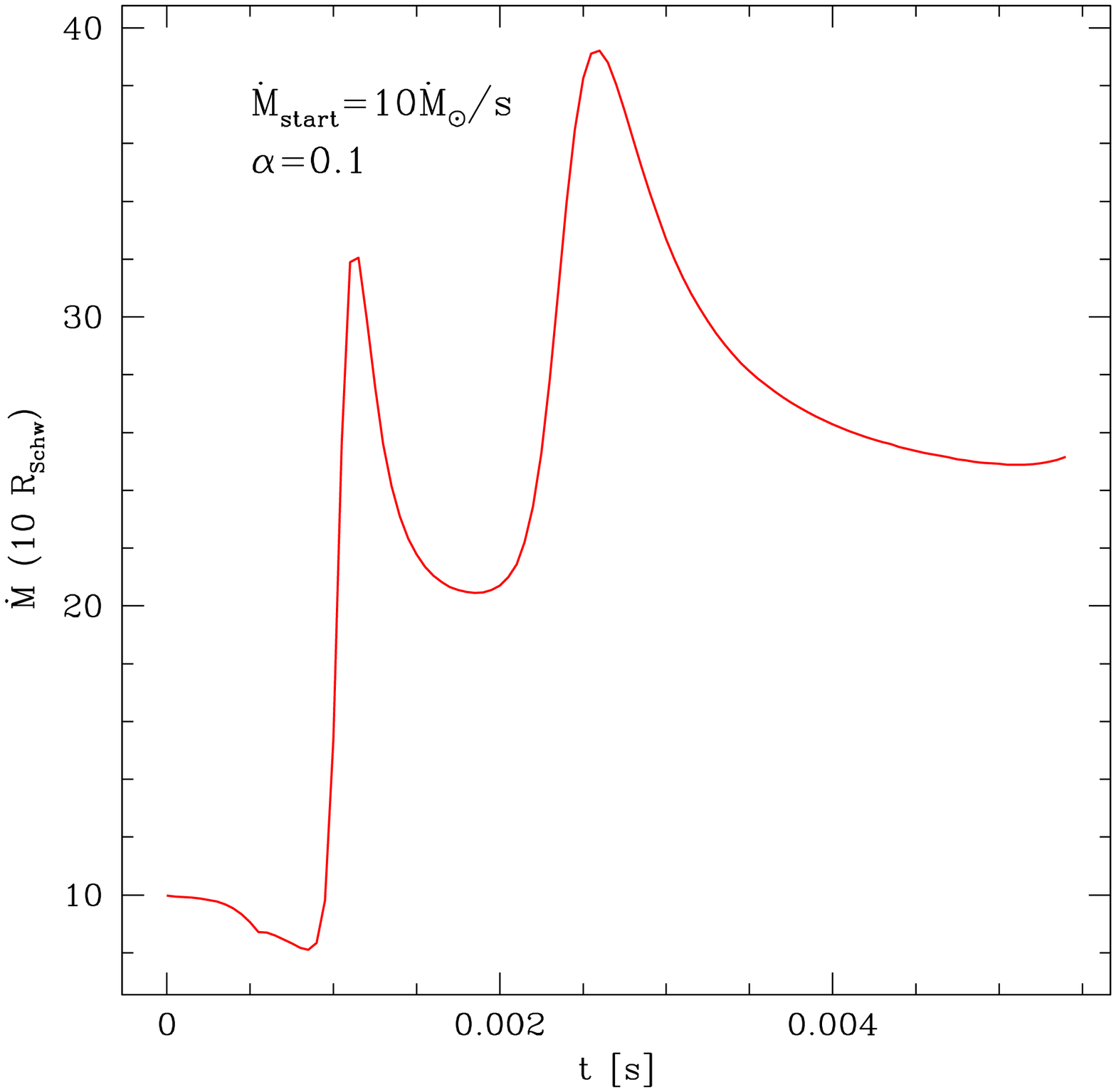}
\caption{Left: stability curves on the 
accretion rate vs. surface density plane, for several chosen radii in the disc:
$3.01 R_{S}$,  $4.06 R_{S}$, $5.27 R_{S}$ and $11.69 R_{S}$.
Right: Fluctuations of the accretion rate at 10 $R_{S}$ due to the disk instability.
}
\end{figure}

\noindent
For $\dot M \approx 10 M_{\odot}/s$ the disk becomes unstable below 10 $R_{S}$ (Fig. 3).  
Here helium is almost completely photodisintegrated while the
electrons become non-degenerate again. For this high accretion rate,
the electron fraction rises inwards 
in the disk. Under these physical conditions the energy balance is affected, leading 
to the thermal and viscous instability (as demonstrated in the stabilitry curves).

The disk instability leads to fluctuations and a variable accretion
rate.  This in turn may cause a variable energy output in the jet, and
hence a variable Lorentz factor $\Gamma$. This
is a fundamental ingredient in order to obtain internal shocks.  This
type of instability was already found in collapsar simulations
(\cite{McFadWoo:99}).

\section{Summary}

 We model the central engine of a GRB with a hyperaccreting disk, cooled by neutrino emission and advection of energy onto the central stellar mass black hole.
The equation of state is determined under the assumption of the $\beta$ 
equilibrium. The chemical composition of matter includes free protons,
neutrons, electrons and positrons, as well as helium nuclei.
Neutrinos are trapped in the disk due to absorption and scattering on nucleons, and for accretion rates   $\dot M \ge 1\,M_\odot$/s the advective cooling in the inner disk parts is relatively more important.
The electron fraction in the disk is much lower than 0.5. The presence of helium nuclei and electron-positron pairs is crucial for the occurrence of an instability in the inner disk.

{\bf Acknowledgments}
This work was supported in part by grant No. PBZ 057/P03/2001 of the
 Polish Commitee for Scientific Research and grant NNSF(10233030).


\bibliographystyle{aipproc}

\bibliography{janiuk_proc}

\end{document}